\documentclass{article}
\usepackage{ijcai17}

\usepackage{times}

\usepackage{graphicx}\usepackage{subfig}
\usepackage{epstopdf}
\usepackage[table]{xcolor}
\usepackage{multirow}
\usepackage{color}
\usepackage{url}   
\usepackage{balance}
\usepackage{fancybox}
\usepackage{booktabs} 
\usepackage{algorithm}\usepackage{algorithmic}
\usepackage{amsmath}\usepackage{amsfonts,amssymb}
\usepackage{bm}
\usepackage{comment}
\usepackage{cases}
\usepackage{threeparttable} 
\usepackage{lscape}
\usepackage[draft]{hyperref}
\hypersetup{
	hidelinks=true,
}

\newcommand{\csharp}{C\nolinebreak\hspace{-0.05em}\raisebox{0.1ex}{\small\bf\# }}
\newcommand{\mysharp}{\nolinebreak\hspace{-0.05em}\raisebox{0.1ex}{\scriptsize\bf\# }}

\begin{document}
	
	\title{DeepAM: Migrate APIs with Multi-modal Sequence to Sequence Learning}	
	
	\author{ 
	Xiaodong Gu$^1$, Hongyu Zhang$^2$, Dongmei Zhang$^3$, and Sunghun Kim$^1$\\
	     $^1$The Hong Kong University of Science and Technology, Hong Kong, China\\
			\{xguaa,hunkim\}@cse.ust.hk\\
		 $^2$The University of Newcastle, Callaghan, Australia\\
		 	hongyu.zhang@newcastle.edu.au\\
	     $^3$Microsoft Research, Beijing, China\\
			dongmeiz@microsoft.com
	}
		
	\maketitle
	
	\begin{abstract}	
		Computer programs written in one language are often required to be ported to other languages to support multiple devices and environments. 
		When programs use language specific APIs~(Application Programming Interfaces), it is very challenging to migrate these APIs to the corresponding APIs written in other languages. 
		Existing approaches mine API mappings from projects that have corresponding versions in two languages. They rely on the sparse availability of bilingual projects, thus producing a limited number of API mappings. 
		In this paper, we propose an intelligent system called \textsc{DeepAM} for automatically mining API mappings from a large-scale code corpus without bilingual projects. The key component of \textsc{DeepAM} is based on the multi-modal sequence to sequence learning architecture that aims to learn joint semantic representations of bilingual API sequences from big source code data.
		Experimental results indicate that DEEPAM significantly increases the accuracy of API mappings as well as the number of API mappings, when compared with the state-of-the-art approaches.
		
	\end{abstract}
	
	
\section{Introduction}\label{s:intro}
	
	Programming language migration is an important task in software development~\cite{mossienko2003automated,hassan2005lightweight,5610429}. A software product is often required to support a variety of devices and environments. This requires developing the software product in one language and manually porting it to other languages. This procedure is tedious and time-consuming. Building automatic code migration tools is desirable to reduce the effort in code migration.	
	
	However, current language migration tools, such as Java2CSharp\footnote{http://j2cstranslator.wiki.sourceforge.net/}, require users to manually define the migration rules between the respective program constructs and the mappings between the corresponding Application Programming Interfaces (APIs) that are used by the software libraries of the two languages. For example, The API \emph{BufferedReader.read} in Java should be mapped to \emph{StreamReader.read} in \csharp. Such a manual procedure is tedious and error-prone. As a result, only a small number of API mappings are produced~\cite{MAM}. 
	
	To reduce manual effort in API migration, several approaches have been proposed to automatically mine API mappings from a software repository~\cite{staminer,TMAP,MAM}. For example, 
	Nguyen et al.~[\citeyear{staminer}] proposed StaMiner that applies statistical machine translation~(SMT)~\cite{phrasesmt} to bilingual projects, namely, projects that are released in multiple programming languages. It first aligns equivalent functions written in two languages that have similar names. Then, it extracts API mappings from the paired functions using the phrase-based SMT model~\cite{phrasesmt}. 
	
	However, existing approaches rely on the sparse availability of bilingual projects. The number of available bilingual projects is often limited due to the high cost of manual code migration.
	For example, we analyzed 11K Java projects on GitHub which were created between 2008 to 2014. Among them, only 15 projects have been manually ported to \csharp versions. 
	Therefore, the number of API mappings produced by existing approaches is rather limited. 
	In addition, given bilingual projects, they need aligning equivalent functions using name similarity heuristics. 
	Only a portion of functions in a bilingual project have similar function names and can be aligned~\cite{MAM}.
	
	In this paper, we propose \textsc{DeepAM}~(Deep API Migration), a novel, deep learning based system to API migration.  
	Without the restriction of using bilingual projects, \textsc{DeepAM} can directly identify equivalent source and target API sequences from a large-scale commented code corpus. 
	The key idea of \textsc{DeepAM} is to learn the semantic representations of both source and target API sequences 
	and identify semantically related API sequences for the migration.
	\textsc{DeepAM} assigns to each API sequence a continuous vector in a high-dimensional semantic space in such a way that API sequences with similar vectors, or ``embeddings'', tend to have similar natural language descriptions. 
	
	In our approach, \textsc{DeepAM} first extracts API sequences~(i.e., sequences of API invocations) from each function in the code corpus. For each API sequence, it assigns a natural language description that is automatically extracted from corresponding code comments.
	With the \textsf{\small$\langle$API sequence, description$\rangle$} pairs, \textsc{DeepAM} applies the sequence-to-sequence learning~\cite{cho2014phrase} to embed each API sequence into a fixed-length vector that reflects the intent in the corresponding natural language description.
	By \emph{jointly embedding} both source and target API sequences into the same space, \textsc{DeepAM} aligns the equivalent source and target API sequences that have the closest embeddings. Finally, the pairs of aligned API sequences are used to extract general API mappings using SMT. 
	

	
	To our knowledge, \textsc{DeepAM} is the first system that applies deep learning techniques to learn the semantic representations of API sequences from a large-scale code corpus. It has the following key characteristics that make it unique:
	\begin{itemize}\setlength{\itemsep}{-0.5\itemsep}
		\vspace{-0.3\baselineskip}
		\item \textbf{Big source code:} \textsc{DeepAM} enables the construction of large-scale bilingual API sequences from big code corpus rather than limited bilingual projects. It learns API semantic representations from 10 million commented code snippets collected over seven years.
		\item \textbf{Deep model:} The multi-modal sequence-to-sequence learning architecture ensures the system can learn deeper semantic features of API sequences than the traditional shallow ones.
	\end{itemize}
	

\vspace{-1\baselineskip}
\section{Related Work}\label{s:related}
    API migration has been investigated by many researchers~\cite{staminer,TMAP,MAM}. 
    Zhong et al.~[\citeyear{MAM}] proposed MAM, a graph based approach to mine API mappings. MAM builds on projects that are released with multiple programming languages. It uses name similarity to align client code of both languages. Then, it detects API mappings between these functions by analyzing their API Transformation Graphs. 
    Nguyen et al.~[\citeyear{staminer}] proposed StaMiner that directly applies statistical machine translation to bilingual projects. 
    
    However, these techniques require the same client code to be available on both the source and the target platforms. Therefore, they rely on the availability of software packages that have been ported manually from the source to the target platform. Furthermore, they use name similarity as a heuristic in their API mapping algorithms. Therefore, they cannot align equivalent API sequences from client code which are similar but independently-developed. 
    
    
    Pandita et al.~[\citeyear{TMAP}] proposed TMAP, which applies the vector space model~\cite{manning2008ir}, an information retrieval technique, to discover likely mappings between APIs. For each source API, it searches target APIs that have similar text descriptions in their API documentation. However, the vector space model they applied is based on the bag-of-words assumption; it cannot identify sentences with semantically related words and with different sequences of words. 
    

    Recently, deep learning technology~\cite{sutskever2014seq2seq,cho2014phrase} has been shown to be highly effective in various domains (e.g., computer vision and natural language processing). 
    Researchers have begun to apply this technology to tackle some software engineering problems. Huo et al. propose a neural model to learn unified features from natural and programming languages for locating buggy source code~\cite{npcnn}. Gu et al. apply sequence-to-sequence learning to generate API sequences from natural language queries~\cite{deepapi}.
    Hence, this study constitutes the first attempt to apply the deep learning approach to migrate APIs between two programming languages.

\section{Method}\label{s:tech}
	Let $\mathcal{A}$=$\{a^{(i)}\}$ denote a set of API sequences where $a^{(i)}$=[$\alpha_1,...,\alpha_{L_a}$] denotes the sequence of API invocations in a function.	
	Suppose we are given a set of source API sequences $\mathcal{A}_S$=$\{a_S^{(i)}\}$ (i.e., API sequences in a source language) and a set of target API sequences $\mathcal{A}_T$=$\{a_T^{(i)}\}$ (i.e., API sequences in a target language).
	Our goal is to find an alignment between $\mathcal{A}_S$ and $\mathcal{A}_T$, namely,
	\vspace{-0.2\baselineskip}
	\begin{equation}
	\small\vspace{-0.4\baselineskip}
	f:\mathcal{A}_S\rightarrow\mathcal{A}_T
	\end{equation} 
	so that each source API sequence $a_S^{(i)}$$\in$$\mathcal{A}_S$ is mapped to an equivalent target API sequence $a_T^{(j)}$$\in$$\mathcal{A}_T$.
	
	Since $\mathcal{A}_S$ and $\mathcal{A}_T$ are heterogeneous, it is difficult to discover the correlation~$f$ directly.
	Our approach is based on the intuition of \textbf{``third party translation''}. That is, although $\mathcal{A}_S$ and $\mathcal{A}_T$ are heterogeneous, in the sense of vocabulary and usage patterns, they can all be mapped to high-level user intents described in natural language. Thus, we can bridge them through their natural language descriptions. For each $a^{(i)}$$\in$$\mathcal{A}$, we assume that there is a corresponding natural language description~$d^{(i)}$=[$w_1,...,w_{L_d}$] represented as a sequence of words.
	
	The idea can be formulated with Joint Embedding(a.k.a., multi-modal embedding)~\cite{xu2015jointly}, a technique to jointly embed/correlate heterogeneous data into a unified vector space so that semantically similar concepts across the two modalities occupy nearby regions of the space~\cite{karpathy2015imagesemantic}.
	In our approach, the joint embedding of $\mathcal{A}_S$ and $\mathcal{A}_T$ can be formulated as:
	\vspace{-0.5\baselineskip}
	\begin{equation}
	\small
	\vspace{-0.4\baselineskip}
	\mathcal{A}_S\xrightarrow[]{\phi} V\xrightarrow[]{\tau} \mathcal{D} \xleftarrow[]{\tau} V \xleftarrow[]{\psi}\mathcal{A}_T
	\end{equation}
	where $V$$\in$$\mathbb{R}^d$ is a common vector space representing the semantics of API sequences; $\phi$$:$$\mathcal{A}_S$$\rightarrow$$V$ is an embedding function to map $\mathcal{A}_S$ into $V$, $\psi$$:$$\mathcal{A}_T$$\rightarrow$$V$ is an embedding function to map $\mathcal{A}_T$ into $V$, $\mathcal{D}$=$\{d_S^{(i)}\}$$\cup$$\{d_T^{(i)}\}$ is the space of natural language descriptions.
	$\tau:V$$\rightarrow$$\mathcal{D}$ is a function to translate from the semantic representations~$V$ to corresponding natural language descriptions~$\mathcal{D}$. 

	\begin{figure} [!tb]
		\setlength{\abovecaptionskip}{3pt}
		\setlength{\belowcaptionskip}{0pt}
		\centering 
		\includegraphics[width=3.5in]{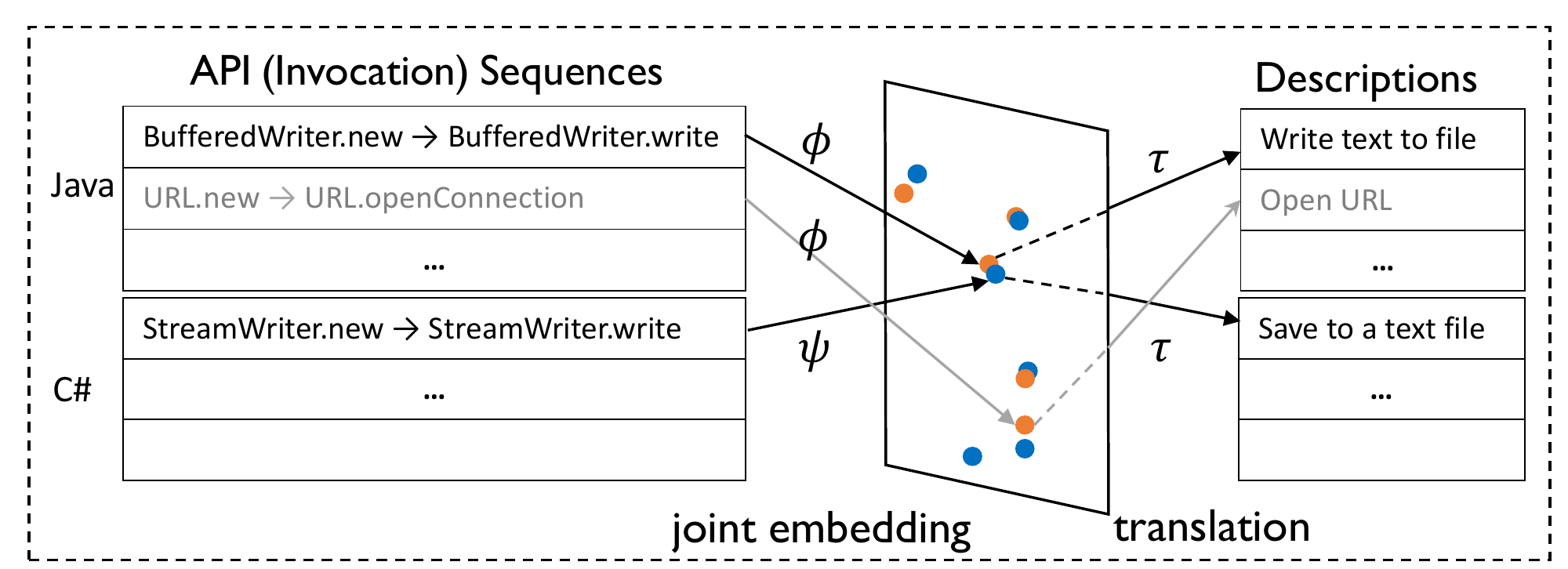} 
		\caption{An Illustration of Joint Semantic Embedding}
		\label{fig:approach:sem_adapt}
		\vspace{-1\baselineskip}
	\end{figure}
	Through joint embedding, $\mathcal{A}_S$ and $\mathcal{A}_T$ can be easily correlated through their semantic vectors  $V_{\mathcal{A}_S}$ and $V_{\mathcal{A}_T}$.
	Figure~\ref{fig:approach:sem_adapt} shows an illustration of joint semantic embedding between Java and \csharp API sequences. We are given a corpus of API sequences (in both Java and \csharp) and the corresponding natural language descriptions. 
	Each API sequence is embedded~(through $\phi$ or $\psi$) and translated~(through $\tau$) to its corresponding description. The yellow and blue points represent embeddings of Java and \csharp APIs respectively. Through traning, the Java API sequence \emph{BufferedWriter.new}$\rightarrow$ \emph{BufferedWriter.write} and the \csharp API sequence \emph{StreamWriter.new}$\rightarrow$\emph{StreamWriter.write} are embedded into a nearby place in order to generate similar corresponding descriptions~\emph{write text to file} and \emph{save to a text file}. 
	Therefore, the two API sequences can be identified as semantically equivalent API sequences.
	
	\subsection{Learning Semantic Representations of API Sequences}\label{ss:tech:apiemb}

	
	In our approach, the semantic embedding function~($\phi$ or $\psi$) and the translation function~$\tau$ are realized using the RNN-based sequence-to-sequence learning framework~\cite{cho2014phrase}.
	The sequence-to-sequence learning is a general framework where the input sequence is embedded to a vector that represents the semantic representation of the input, and the semantic vector is then used to generate the target sequence. The model that embeds the sequence to a vector (i.e., $\phi$ or $\psi$) is called ``encoder'', and the model that generates the target sequence(i.e., $\tau$) is called ``decoder''.
	
\begin{figure} [!tb]
	\setlength{\abovecaptionskip}{0pt}
	\setlength{\belowcaptionskip}{0pt}
	\centering 
	\includegraphics[width=3.2in]{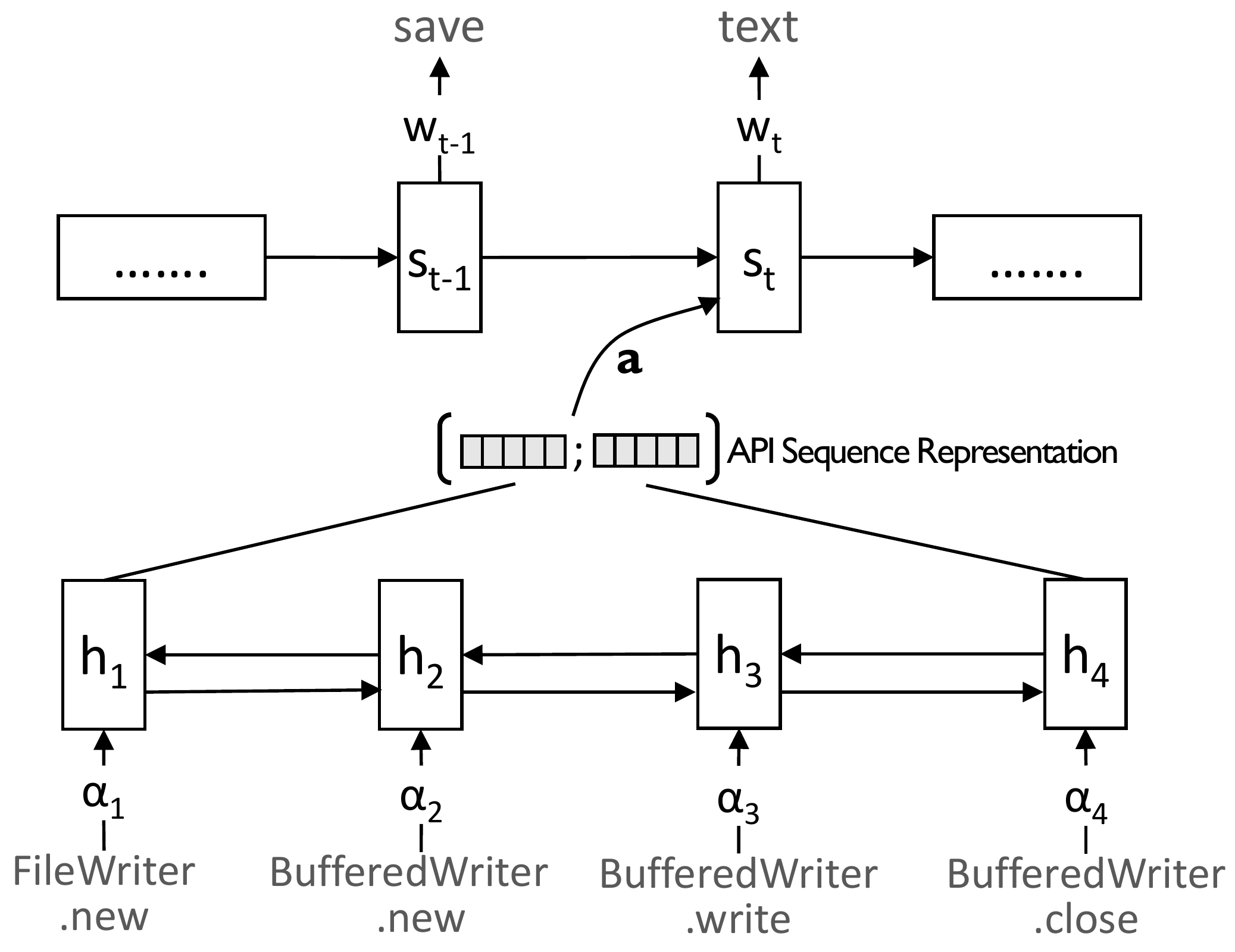} 
	\caption{The sequence-to-sequence learning framework for API Semantic Embedding. A bidirectional RNN is used to concatenate the forward and backward hidden states as the semantic representations of API sequences}
	\label{fig:approach:rnnencdec}
	\vspace{-1\baselineskip}
	\end{figure}

	The framework of the sequence-to-sequence model applied	to API semantic embedding is illustrated in Figure~\ref{fig:approach:rnnencdec}. Given a set of \textsf{\small$\langle$API sequence, description$\rangle$} pairs~$\{\langle a^{(i)},d^{(i)}\rangle\}$,
	The encoder (a bi-directional recurrent neural network~\cite{mikolov2010rnnlm}) converts each API sequence~$a$=[$\alpha_1, ...,\alpha_{L_a}$], to a fixed-length vector~$\bm{a}$ using the following equations iteratively from $t$ = 1 to $L_d$:
	\vspace{-0.2\baselineskip}
	\begin{align}
	\small
	\label{eq:encdec:h} \bm{h}_t&=tanh(\bm{W}_{enc}[\bm{h}_{t-1};\bm{\alpha}_t]+\bm{b}_{enc})\\
	\vspace{-0.9\baselineskip}
	\label{eq:encdec:c} \bm{a}&=\bm{h}_{L_a}
	\end{align}
	where $\bm{h}_t (t$=$1,...,L_a)$ represents the hidden states of the RNN at each potion~$t$ of the input; [$a$;$b$] represents the concatenation of two vectors, $\bm{W}_{enc}$ and $\bm{b}_{enc}$ are trainable parameters in the RNN, $tanh$ is the activation function.

	The decoder then uses the encoded vector to generate the corresponding natural language description~$d$ by sequentially predicting a word~$w_t$ conditioned on the vector~$\bm{a}$ as well as previous words $w_1,...,w_{t-1}$.
	\vspace{-0.5\baselineskip}
	\begin{align}
	\scriptsize
	\vspace{-0.1\baselineskip}
	\label{eq:encdec:pred1} Pr(d)=&\prod_{t=1}^{L_d} p(w_t|{w_1,...,w_{t-1}},\bm{a})\\
				p(w_t|w_1,...,&w_{t-1},\bm{a})=\mathrm{softmax}(\bm{W}_{dec}^o\bm{s}_t+\bm{b}_{dec}^o)\\
	\label{eq:encdec:pred3} \bm{s}_t=tanh&(\bm{W}_{dec}^s[\bm{s}_{t-1};\bm{w}_{t-1};\bm{a}]+\bm{b}_{dec}^s)
	\end{align} 
	where $\bm{s}_t(t$=$1,...,L_d)$ represents the hidden states of the RNN at each potion~$t$ of the output; $\bm{W}_{dec}^o$, $\bm{b}_{dec}^o$, $\bm{W}_{dec}^s$ and $\bm{b}_{dec}^s$ are trainable parameters in the decoder RNN.
	
	Both the encoder and decoder RNNs are implemented as a bidirectional gated recurrent neural network~(GRU)~\cite{cho2014phrase} which is a widely used implementation of RNN
	. Both GRUs have two hidden layers, each with 1000 hidden units. 

	\subsection{Joint Semantic Embedding for Aligning \\ Equivalent API Sequences}\label{ss:tech:apijointemb}
	
	For \textit{joint embedding}, we train the sequence-to-sequence model on both $\{\langle a_S^{(i)},d_S^{(i)}\rangle\}$ and $\{\langle a_T^{(i)},d_T^{(i)}\rangle\}$ to minimize the following objective function:
	\vspace{-0.4\baselineskip}
	\begin{equation}\label{eq:rnnenecdec:obj}
	\small
	\begin{split}
	\mathcal{L}(\theta) =&-\frac{1}{N_S}\sum_{i=1}^{N_S}\sum_{t=1}^{L_d}\log p_\theta(w_S^{(it)}|a_S^{(i)})\\
	 &-\frac{1}{N_T}\sum_{i=1}^{N_T}\sum_{t=1}^{L_d}\log p_\theta(w_T^{(it)}|a_T^{(i)})
	\end{split}
	\vspace{-0.4\baselineskip}
	\end{equation}
	where $N_S$ and $N_T$ are the total number of source and target training instances, respectively. $L_d$ is the length of each natural language sentence. 
	$\theta$ denotes model parameters, while $p_{\theta}(w^{(it)}|a^{(i)})$ (derived from Equation~\ref{eq:encdec:h} to \ref{eq:encdec:pred3}) denotes the likelihood of generating the $t$-th target word given the API sequence~$a^{(i)}$ according to the model parameters~$\theta$.
	
	After training, each API sequence $a$=[$\alpha_1,...,\alpha_{L_a}]$ is embedded to a vector $\bm{a}$ that reflects developer's high-level intent.
	We identify equivalent source and target API sequences as those having close semantic vectors.
	
\section{Implementation}\label{s:approach}
\begin{figure*} [!tb]
		\setlength{\abovecaptionskip}{3pt}
		\setlength{\belowcaptionskip}{0pt}
		\centering 
		\includegraphics[width=6.8in]{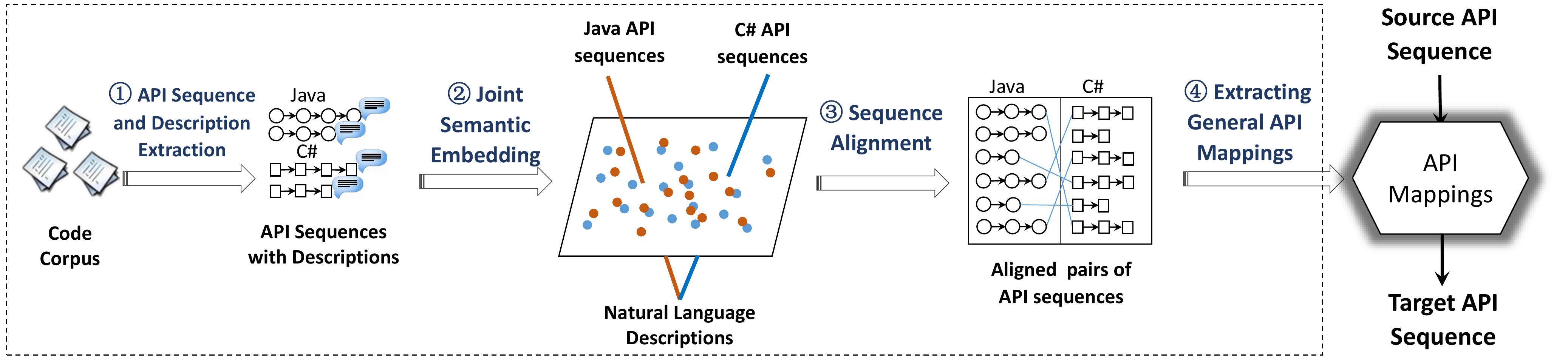} 
		\caption{ The Overall Workflow of \textsc{DeepAM}}
		\label{fig:approach:framework}
		\vspace{-1\baselineskip}
	\end{figure*}

	In this section, we describe the detailed implementation of \textsc{DeepAM}, a deep-learning based system we propose to migrate API usage sequences. Figure~\ref{fig:approach:framework} shows the overall workflow of \textsc{DeepAM}. It includes four main steps. We first prepare a large-scale corpus of \textsf{\small$\langle$API sequence, description$\rangle$} pairs 
	for both Java and \csharp (Step~1). The pairs of both languages are jointly embedded by the sequence-to-sequence model as described in Section~\ref{ss:tech:apijointemb} (Step~2). Then, we identify related Java and \csharp API sequences according to their semantic vectors (Step~3). Finally, a statistical machine translation component is used to extract general API mappings from the aligned bilingual API sequences (Step~4). 
	
	In theory, our system could migrate APIs between any programming languages. In this paper we limit our scope to the Java-to-\csharp migration. The details of each step are explained in the following sections.		
	
	\subsection{Gathering a Large-scale API Sequence-to- Description Corpus}\label{ss:approach:data}
	We first construct a large-scale database that contains \textsf{\small$\langle$API sequence, description$\rangle$} pairs for training the model. We download Java and \csharp projects created from 2008 to 2014 from GitHub\footnote{http://github.com}. To remove toy or experimental programs, we only select the projects with at least one star. In total, we collected 442,928 Java projects and 182,313 \csharp projects from GitHub. 
	
	Having collected the code corpus, we extract API sequences and corresponding natural language descriptions: 
	we parse source code files into ASTs (Abstract Syntax Trees) using Eclipse's JDT compiler\footnote{http://www.eclipse.org/jdt} for Java projects, and Roslyn\footnote{https://roslyn.codeplex.com/} for \csharp projects.
	Then, we extract the API sequence from individual functions using the same approach in \cite{deepapi}. 
	
\begin{figure} [!tb]
		\setlength{\abovecaptionskip}{3pt}
		\setlength{\belowcaptionskip}{0pt}
		\centering 
		\includegraphics[width=3.2in]{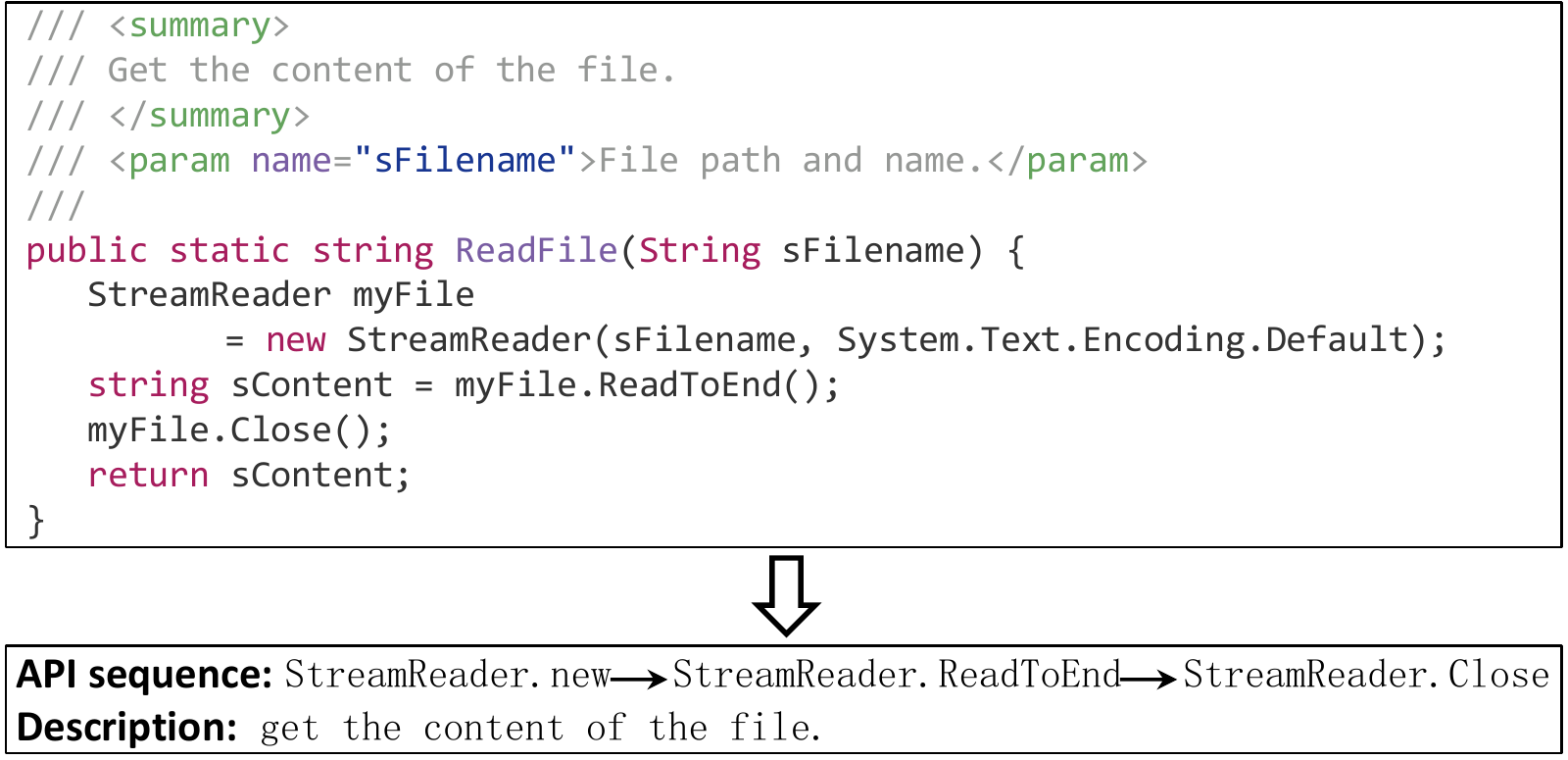}
		\caption{\small An example of extracting an API sequence and its description from a C\# function \emph{TextFile.ReadFile}$^{7}$}
		\label{fig:approach:annotation}
		\vspace{-1\baselineskip}
	\end{figure}

	To obtain natural language descriptions for the extracted API sequences, we extract function-level code summaries from code comments. In both Java and \csharp, it is the first sentence of a \emph{documentation comment}\footnote{A \emph{documentation comment} in Java starts with `/**' and ends with `*/'. A \emph{documentation comment} in \csharp starts with a ``$<$summary$>$'' tag and ends with a ``$<$/summary$>$'' tag.} for a function. According to the Javadoc guidance\footnote{http://www.oracle.com/technetwork/articles/java/index-137868.html}, the first sentence of a documentation comment is used as a short summary of a function.
	Figure~\ref{fig:approach:annotation} shows an example of \emph{documentation comments} for 
	a \csharp function \emph{TextFile.ReadFile}\footnote{https://github.com/virtualmarc/gitlab-ci-runner-win/blob/master/gitlab-ci-runner/helper/TextFile.cs} in the Gitlab CI project. 
	
	
	Finally, we obtain a database consisting of 9,880,169 \textsf{\small$\langle$API sequence, description$\rangle$} pairs, including 5,271,526 Java pairs and 4,608,643 \csharp pairs. 
	
	\subsection{Model Training}\label{ss:approach:seqemb}
	We train the sequence-to-sequence model on the collected \textsf{\small$\langle$API sequence, description$\rangle$} pairs of both Java and \csharp. 
	The model is trained using the mini-batch stochastic gradient descent algorithm~(SGD)~\cite{sgd} together with Adadelta \cite{zeiler2012adadelta}. 
	We set the batch size as 200. Each batch is constituted with 100 Java pairs and 100 \csharp pairs that are randomly selected from corresponding datasets. The vocabulary sizes of both APIs and natural language descriptions are set to 10,000.
	The maximum sequence lengths~$L_a$ and $L_d$ are both set as 30. Sequences that exceed the maximum lengths will be excluded for training. 
	
	After training, we feed in the encoder with all API sequences and obtain corresponding semantic vectors from the last hidden layer of encoder.
	
	\subsection{API Sequence Alignment}\label{ss:approach:seqalign}
	After embedding all API sequences, we build pairs of equivalent Java and \csharp API sequences according to their semantic vectors. For each Java API sequence, we find the most related \csharp API sequence to align with by selecting the \csharp API sequence that has the most similar vector representation. We measure the similarity between the vectors of two API sequences using the cosine similarity, which is defined as: 
	\vspace{-0.6\baselineskip}
	\begin{equation}
	\small
	\mathrm{similarity}(\bm{a}_s,\bm{a}_t)=\frac{\bm{a}_s\cdot\bm{a}_t}{\parallel\bm{a}_s\parallel\parallel\bm{a}_t\parallel}
	\vspace{-0.5\baselineskip}
	\end{equation}
	where $\bm{a}_s$ and $\bm{a}_t$ are vectors of source and target API sequences. The higher the similarity, the more related the source and target API sequences are to each other. 
	
	Finally, we obtain a database consisting of aligned pairs of Java and \csharp API sequences.	
	\begin{table*}[tb]
	\setlength{\abovecaptionskip}{0pt}
	\setlength{\belowcaptionskip}{0pt}
	\centering
	\scriptsize
	\caption{Accuracy of 1-to-1 API mappings mined by \textsc{DeepAM} and StaMiner (\%)}
	\begin{tabular}{p{1.1cm}|@{}c@{}|@{}c@{}|@{}c@{}|@{}c@{}|@{}c@{}|@{}c@{}|@{}c@{}|@{}c@{}|@{}c@{}|@{}c@{}|@{}c@{}|@{}c@{}} 
		\toprule[1pt]
		\multirow{3}{*}{Package} & \multicolumn{6}{c|}{Class Migration} & \multicolumn{6}{c}{Method Migration} \\ \cline{2-13}
		&\multicolumn{2}{c|}{Precision}  & \multicolumn{2}{c|}{Recall} 	& \multicolumn{2}{c|}{F-score}  &\multicolumn{2}{c|}{Precision}  & \multicolumn{2}{c|}{Recall} 	& \multicolumn{2}{c}{F-score}  \\ \cline{2-13}
		&	\;StaMiner\;& \;DeepAM\;	& \;StaMiner\;  & \;DeepAM\;  	& \; StaMiner\; &\; DeepAM\;	&	\;StaMiner\;& \;DeepAM\;	& \;StaMiner\;  & \;DeepAM\;  	& \; StaMiner\; &\; DeepAM\;	\\ \hline
		java.io				&  		70.0\%	&	80.0\%	&	63.6\% 		&  	75.0\%		&	66.6\%		&   72.7\% 	&  		70.0\%	&		66.7\%	&	64.0\% 		&  	87.5\%		&	66.9\%		&   75.2\% 		\\ 
		java.lang			& 		82.5\% 	& 	80.0\%	&	76.7\% 		& 	81.3\%		& 	79.5\% 		&	80.7\%  & 		86.7\% 	& 		83.7\%	&	76.5\% 		& 	87.2\%		& 	81.3\% 		&	85.4\%\\ 
		java.math			&  		50.0\%	&	66.7\%	&	50.0\% 		&	66.7\%		& 	50.0\%		&	66.7\%  &  		66.7\%	&		66.7\%	&	66.7\% 		&	66.7\%		& 	66.7\%		&	66.7\% \\   
		java.net 			&		100.0\%	&	100.0\%	&	50.0\%		&	100.0\%		&	66.7\%		&	100.0\% &		100.0\%	&		100.0\%	&	33.3\%		&	100.0\%		&	50.0\%		&	100.0\%	\\
		java.sql 			&		100.0\%	&	100.0\%	&	50.0\%		&	100.0\%		&	66.7\%		&   100.0\% &		100.0\%	&		50.0\%	&	50.0\%		&	66.7\%		&	66.7\%		&   57.2\% \\
		java.util 			&		64.7\%	&	69.6\%	&	71.0\%		&	72.7\%		&	67.7\%		&   71.1\%  &		63.0\%	&	64.3\%		&	54.8\%		&	85.7\%		&	58.6\%		&   73.5\%\\ \hline
		All					& 		77.9\%	&\bf82.7\%	&	60.2\%		&\bf82.6\%		&	66.2\%		&\bf 81.9\% & 		81.1\%	&	71.9\%		&	57.6\%		&\bf82.3\%		&	65.0\%		& \bf 76.3\% \\ 
		\bottomrule[1pt] 
	\end{tabular}
	\label{tab:result:accuracy:single}
	\vspace{-1.5\baselineskip}
	\end{table*}	
	\subsection{Extracting General API Mappings}	
	The aligned pairs of API sequences may be project-specific. However, automated code migration tools such as Java2CSharp require commonly used API mappings.
	To obtain more general API mappings, we summarize mappings that have high co-occurrence probabilities in the aligned pairs of API sequences. To do so, we apply an SMT technique named \emph{phrase-based model}~\cite{phrasesmt} to the pairs of aligned API sequences. The phrase-based model was originally designed to extract phrase-to-phrase translation mappings from bilingual sentences. In our system, the phrase model summarizes pairs of API phrases, namely, subsequences of APIs that frequently co-occur in the aligned pairs of API sequences. 
	For each phrase pair, it assigns a score defined as the translation probability
	$p(t|s)$=$\mathrm{count}(s,t)/(\mathrm{count}(s)$+$1)$, where $\mathrm{count}(s,t)$ is the number of mapping occurrences $s$$\rightarrow$$t$, and $\mathrm{count}(s)$ is the number of all occurrences of the subsequence~$s$. 
	Finally, we select pairs whose translation probabilities are greater than a threshold as the final API mappings.   
	We set the threshold to 0.5 as in StaMiner~\cite{staminer}

\section{Experimental Results}\label{s:eval}

	\subsection{Accuracy in Mining API Mappings}
	We first evaluate how accurate \textsc{DeepAM} performs in mining API mappings. 
	We focus on 1-to-1 API mappings that are currently used by many code migration tools such as Java2Csharp. 
	We compare the 1-to-1 API mappings mined by \textsc{DeepAM} (Section 4) with a ground truth set of manually written API mappings provided by Java2CSharp. 
	\\\textbf{Metric}
	We use the F-score to measure the accuracy. It is defined as:
		$F$=$2PR/(P$+$R)$
		where
		$P$=$\frac{TP}{TP+FP}$
		and
		$R$=$\frac{TP}{TP+FN}$.
	\emph{TP} is true positive, namely, the number of API mappings that are both in \textsc{DeepAM} results and in the ground truth set. \emph{FP} is false positive which represents the number of resulting mappings that are not in the ground truth set. 
	\emph{FN} is false negative, which represents the number of mappings that are in the ground truth set but not in the results.
	\\\textbf{Baselines} We compare \textsc{DeepAM} with StaMiner~\cite{staminer} and TMAP~\cite{TMAP}.
	StaMiner is a state-of-the-art API migration approach that directly utilizes statistical machine translation on bilingual projects. 
	TMAP~\cite{TMAP} is an API migration approach using information retrieval techniques. It aligns Java and \csharp APIs by searching similar descriptions in API documentation. For easy comparison, we use the same configuration as in TMAP~\cite{TMAP}. 
	We manually examine the numbers of correctly mined API mappings on several Java SDK classes and make a direct comparison with the TMAP's results presented in their paper.
	\\\textbf{Results}		
	Table~\ref{tab:result:accuracy:single} shows the accuracy of both \textsc{DeepAM} and StaMiner. We evaluate the accuracy of mappings for both API classes and API methods.
	The results show that \textsc{DeepAM} is able to mine more correct API mappings. It achieves average recalls of 82.6\% and 82.3\% for class and method migrations respectively, which are significantly greater than StaMiner~(60.2\% and 57.6\%). 
	The average precisions of \textsc{DeepAM} are 82.7\% and 71.9\%, slightly less than but similar to StaMiner~(77.9\% and 81.1\%). 
	Overall, \textsc{DeepAM} performs better than StaMiner, with average F-measures of 81.9\% and 76.3\% compared to StaMiner 's~(66.2\% and 65.0\%).  
	
	Table~\ref{tab:result:comp:tmap} shows the number of correctly mined API mappings by TMAP and \textsc{DeepAPI}. The column \mysharp \textsf{\small Methods} lists the total numbers of API methods for each class. As shown in the results, \textsc{DeepAM} can mine many more correct API mappings than TMAP, which is based on text similarity matching. 
	
	The results indicate that without the restriction of a few bilingual projects, \textsc{DeepAM} yields many more correct API mappings.
		\begin{table}[tb]
		\setlength{\abovecaptionskip}{0pt}
		\setlength{\belowcaptionskip}{0pt}
		\centering
		\scriptsize
		\caption{\small Number of correct API mappings mined by \textsc{DeepAM} and TMAP}
			\begin{tabular}{p{2.4cm}|c|p{0.8cm}|p{0.8cm}} 
				\toprule[1pt]
				\multirow{2}{*}{Class}& \mysharp &\multicolumn{2}{c}{\mysharp API mappings} \\ \cline{3-4}
				&Methods&	TMAP		& \textsc{DeepAM} 	  \\ \hline
				java.io.File			& 54 	&  26			& 43	 		\\ 
				java.io.Reader			& 10 	&	6 		&  	8	\\ 
				java.io.Writer			& 10 	&	10 		&  	7	\\ 
				java.util.Calendar		& 47	&	5 		&  	20	\\ 
				java.util.Iterator		& 3	 	&	1 		&  	3	\\ 
				java.util.HashMap		& 17	&	5 		&  	14	\\ 
				java.util.ArrayList		& 28	&	15 		&  	26	\\ 
				java.sql.Connection		& 52	&	13 		&  	23	\\ 
				java.sql.ResultSet		& 187	&	31 		&  	33	\\ 
				java.sql.Statement		& 42	&	5 		&  	15	\\ \hline
				All						&	450	&	117		&	192	\\ 
				\bottomrule[1pt] 
			\end{tabular}
		\label{tab:result:comp:tmap}
		\vspace{-1\baselineskip}
	\end{table}	
	\begin{table}[tb]
	\setlength{\abovecaptionskip}{0pt}
	\setlength{\belowcaptionskip}{0pt}
	\centering
	\scriptsize
	\caption{\small Number of API Mappings Mined by \textsc{DeepAM} and StaMiner}
	\begin{tabular}{c|@{}c@{}|@{}c@{}|@{}c@{}|@{}c@{}|@{}c@{}|@{}c@{}|@{}c} 
		\toprule[1pt]
		\multirow{2}{*}{Tool}	&		\multicolumn{5}{c|}{{\mysharp}API Mapping Rules by Sequence Length}   &  \multirow{2}{*}{Corr.}& \multirow{2}{*}{EDR}	  	 \\ \cline{2-6}
		&	1& 2-3 & 4-7 & 8+ & Total &		&\\ \hline
		StaMiner		& \;50,992\; & 31,754 & 14,370   & 3,708	& 100,825	&	\;87.1\%\; & \;7.3\%\;		\\ \hline 
		\textsc{DeepAM}	 &\;35,973\; &\;218,957\;&  \; 328,290\;	&\; 225,268\;&\;808,488\;	& \;88.2\%\; & \;8.2\%\;\\   
		\bottomrule[1pt] 
	\end{tabular}
	\label{tab:result:scale}
	\vspace{-1\baselineskip}
	\end{table}
	\begin{table*}[tb]
	\setlength{\abovecaptionskip}{0pt}
	\setlength{\belowcaptionskip}{0pt}
	\scriptsize
	\caption{Examples of Mined API Mappings}
	\begin{tabular}{p{2.55cm}|p{7.4cm}|p{6.8cm}@{}} 
		\toprule[1pt]
		\bf Task & \bf Java API Sequence & \bf Migrated C\mysharp API sequence by \textsc{DeepAM}\\ 	\toprule[1pt]
		parse datetime from string   & SimpleDateFormat.new SimpleDateFormat.parse&DateTimeFormatInfo.new DateTime.parseExact DateTime.parse\\\hline
		open a url&  URL.new URL.openConnection& WebRequest.create Uri.new HttpWebRequest.getRequestStream\\\hline
		get files in folder&   File.new File.list File.new File.isDirectory & DirectoryInfo.new DirectoryInfo.getDirectories\\\hline
		generate md5 hash code & MessageDigest.getInstance MessageDigest.update MessageDigest.digest& MD5.create UTF8Encoding.new UTF8Encoding.getBytes MD5.computeHash\\\hline
		execute sql statement &Connection.prepareStatement PreparedStatement.execute& SqlConnection.open SqlCommand.new SqlCommand.executeReader\\\hline
		create directory&File.new File.exists File.createNewFile& FileInfo.new Directory.exists Directory.createDirectory\\\hline
		read file &System.getProperty FileInputStream.new InputStreamReader.new BufferedReader.new BufferedReader.read BufferedReader.close&FileInfo.new StreamReader.new StreamReader.read StreamReader.close \\\hline
		create socket&InetSocketAddress.new ServerSocket.new ServerSocket.bind ServerSocket.close&Socket.new IPEndPoint.new Socket.bind Socket.close\\\hline
		download file from url & URL.new URL.openConnection URLConnection.getInputStream BufferedInputStream.new&WebRequest.create HttpWebRequest.getResponse HttpWebResponse.getResponseStream StreamReader.new\\\hline
		save an image to a file & BufferedImage.new Color.new Color.getRGB BufferedImage.setRGB String.endsWith File.new ImageIO.write&Bitmap.new Color.new Color.fromArgb Bitmap.setPixel Bitmap.save \\\hline
		parse xml & DocumentBuilderFactory.newInstance DocumentBuilderFactory.newDocumentBuilder DocumentBuilder.parse &XDocument.load HttpUtility.htmlEncode XDocument.parse\\\hline
		play audio& AudioSystem.getClip File.new AudioSystem.getAudioInputStream Clip.open Clip.start Clip.isRunning Thread.sleep Clip.close &SoundPlayer.new SoundPlayer.play Thread.sleep SoundPlayer.stop\\ 
		\bottomrule[1pt]
	\end{tabular}
	\label{tab:eval:extrinsic}
	\vspace{-1\baselineskip}
	\end{table*}

	\subsection{The Scale of Mined API Mappings}	
	We also evaluate the scalability of \textsc{DeepAM} on mining API mappings: 
	we compare the number of API mappings minend by \textsc{DeepAM} and StaMiner~\cite{staminer} with respect to sequence lengths. We can make this comparison because both DeepAM and StaMiner support sequence-to-sequence mapping. 
	We also consider the quality of mined API mappings in the comparison.
	We use \emph{correctness} and \emph{edit distance ratio} (EDR) to measure the quality as used in~\cite{staminer}.
	The \emph{correctness} is defined as the percentage of correct API sequences of all the migrated results. 
	The \emph{EDR} is defined as the ratio of elements that a user must delete/add in order to transform a result into a correct one.
	\vspace{0.2\baselineskip}
	$\textrm{EDR}$=$\frac{\sum_{\mathrm{pairs}}\mathrm{EditDist}(s_R,s_T)}{\sum_{\mathrm{pairs}}\mathrm{length}(s_T)}$
	, where $\mathrm{EditDist}$ measures the edit distance between the ground truth sequence~$s_R$ and the result sequence $s_T$; $\mathrm{length}$($s_T$) is the number of symbols in~$s_T$.
	The value of EDR ranges from 0 to 100\%. The smaller the better. 
	\\\textbf{Results}
	Table~\ref{tab:result:scale} shows the number of API mappings produced by \textsc{DeepAM} and StaMiner. Each column within \mysharp \textsf{\small API Mappings by Sequence Length} shows the number of mined API mappings within a specific range of length: one~(column 1), two or three~(2-3), four to seven~(4-7), and eight or more~(8+).
	As we can see, \textsc{DeepAM} produces many more API mappings than StaMiner, with comparable quality. The total number of mappings mined by \textsc{DeepAM} is 808,488, which is significantly greater than that of StaMiner~(100,825). In particular, DeepAM produces more mappings for long API sequences. 
	The quality of mappings by \textsc{DeepAM} is comparable to that by StaMiner. The correctness of \textsc{DeepAM} is 88.2\%, which is slightly greater than that of StaMiner (87.1\%). However, mappings produced by \textsc{DeepAM} need slightly more error correlations than StaMiner. 

	Overall, the results indicate that \textsc{DeepAM} significantly increases the number of API mappings than StaMiner, with comparable quality.	
	These results are expected because \textsc{DeepAM} does not rely upon bilingual projects, therefore significantly increasing the size of available training corpus. 
	
	Table~\ref{tab:eval:extrinsic} shows some concrete examples of API mappings. We selected 12 programming tasks that are commonly used in the literature~\cite{codehow,deepapi}. The results show that \textsc{DeepAM} can successfully migrate API sequences for these tasks. 
	\textsc{DeepAM} also performs well in longer API sequences such as \emph{copy file} and \emph{play audio}.

	\begin{table}[tb]
		\setlength{\abovecaptionskip}{0pt}
		\setlength{\belowcaptionskip}{0pt}
		\centering
		\scriptsize
		\caption{Accuracy of API pair alignment by \textsc{DeepAM} and IR-based technique}
		\begin{tabular}{p{1.8cm}|p{1.4cm}|p{1.4cm}|p{0.8cm}} 
			\toprule[1pt]
			Tool			&    Java version	&	C\mysharp version &  	Average		 \\ \hline
			IR				&       37.4\%      &       44.1\%      &   40.8\%       \\ \hline
			\textsc{DeepAM}	&  		60.2\%		&   	84.6\%		& 	72.4\%		\\
			\bottomrule[1pt] 
		\end{tabular}
		\label{tab:result:apiemb}
		\vspace{-1.5\baselineskip}
	\end{table}

	\subsection{Effectiveness of Multi-modal API Sequence Embedding}\label{ss:eval:rq3}

		As the most distinctive feature of our approach is the multi-modal semantic embedding of API sequences, we also evaluate \textsc{DeepAM}'s effectiveness in embedding API sequences, namely, whether the joint embedding is effective on API sequence alignment.
		As described in Section~\ref{ss:approach:seqalign}, we apply the semantic embedding and sequence alignment on raw API sequences, and obtain a database of semantically related Java and \csharp API sequences. 
		We randomly select 500 aligned pairs of Java and \csharp API sequences from the database and manually examine whether each pair is indeed related. We calculate the ratio of related pairs of the 500 sampled pairs.	
		\\\textbf{Baseline}
		We compare our results with an IR based approach. This approach aligns API sequences by directly matching corresponding descriptions using text similarities~(e.g., the vector space model)~\cite{manning2008ir}.
		We implement it using Lucene\footnote{https://lucene.apache.org/}. For each Java API sequence, we search the \csharp API sequence whose description is most similar to the description of the Java API sequence, and vice versa. We randomly select 500 aligned pairs from the results and manually examine the ratio of correctly aligned pairs.
		\\\textbf{Results}
		Table~\ref{tab:result:apiemb} shows the performance of sequence alignment. 
		The column \textsf{\small Java version} shows the ratio of Java API sequences which are correctly aligned to \csharp API sequences. Likewise, the \csharp \textsf{\small version} column shows the ratio of \csharp API sequences that are correctly aligned to Java API sequences. The results show that the joint embedding is effective for the API sequence alignment. 
		The ratio of successful alignments is 72.4\%, which significantly outperforms the IR based approach (average accuracy is 40.8\%).			
		The results indicate that the deep learning model is more effective in learning semantics of API sequences than traditional shallow models such as the vector space model.

\section{Conclusion}\label{s:conclusion}
	In this paper, we propose a deep learning based approach to the migration of APIs. 
	Without the restriction of using bilingual projects, our approach can align equivalent API sequences from a large-scale commented code corpus through multi-modal sequence-to-sequence learning. 
	Our experimental results have shown that the proposed approach significantly increases the accuracy and scale of API mappings the state-of-the-art approaches can achieve.
	Our work demonstrates the effectiveness of deep learning in API migration and is one step towards automatic code migration.
				
				

\balance
\bibliographystyle{named}
\bibliography{references} 
				
\end{document}